\documentclass[aps,amsmath,twocolumn,amssymb,floatfixng,showpacs,
superscriptaddress,footinbib]{revtex4-1}
\pdfoutput=1
\usepackage[dvips]{graphics}
\usepackage{bm}
\usepackage{placeins}
\usepackage{float}

\usepackage{epsfig}
\usepackage{enumerate}
\usepackage{subfigure}
\usepackage{array}
\usepackage{booktabs}
\usepackage{xcolor}
\usepackage{amsmath}
\usepackage{color}
\usepackage{braket}
\usepackage{graphicx}
\usepackage{float}
\usepackage[colorlinks=true,linktoc=page,linkcolor=red,citecolor=blue,urlcolor=magenta]{hyperref}
\usepackage{soul}

\newcommand\bea{\begin{eqnarray}}
\newcommand\eea{\end{eqnarray}}
\newcommand\beq{\begin{equation}}  
\newcommand\eeq{\end{equation}}

\begin{document} 

\title{Finite-size scaling properties of classical random walk on various two-dimensional lattices} 
 \author{Nimish Sharma}
%\email{XX.in}
\affiliation{Department of Physics, BITS Pilani-Pilani Campus, Rajasthan, 333031, India}
\author{Tanay Nag}
\email{tanay.nag@hyderabad.bits-pilani.ac.in}
\affiliation{Department of Physics, BITS Pilani-Hyderabad Campus, Telangana 500078, India}

%-------------------------------------------------------------------------------------------------------------------------------------------------
\begin{abstract}
We consider various two-dimensional lattices such as square, Kagome, Lieb, honeycomb, dice lattices of finite extent, to study the effect of  lattice profile in terms of the number of nearest neighbour and connectivity patterns on the classical random walk in the unbiased scenario.  We find that the  standard deviation of distance travelled by the walker i.e., root mean square displacement of the walker is insensitive to the non-uniformity of the 
lattice profile leading to diffusive transport even in the finite size lattices. We next study the scaling complexity of the entire closed curve traced by the walker while investigating the bulk and boundary fractal dimension namely mass and hull dimensions.
\textcolor{black}{ Our study indicates that the mass fractal dimension varies within a window $1.50\pm 0.03$ for all finite-size lattices. A weak ordering within the above window, correlated with the average coordination number, is observed, while Lieb and square lattices yielding the
minimum and maximum values, respectively. However, confidence intervals
reveal substantial statistical overlap for several lattice pairs even though the lattice profiles vary as far as  the average number of connecting bonds and directionality of bonds are concerned.} 
We also study the scaling complexity of the circumference of the closed curve traced by the walker while investigating the hull dimension.  
We find similar trend for hull fractal dimension as well and that was found to within the window $1.37\pm 0.03$ for finite-size lattices. \textcolor{black}{Within the above window, the ordering remains 
qualitatively unaltered as compared to mass dimension while the confidence interval rectifies the order quantitatively. The 
square lattice  clearly exhibits the upper bound for hull fractal dimension and the remaining lattices show  extensive statistical overlap within the above window. }
We exhibit a tendency of the mass and hull fractal dimension to reach their thermodynamic values given by Brownian motion when we allow more number of steps within the finite size of the lattice, \textcolor{black}{as confirmed by a data collapse analysis}.  
Therefore, our study uncovers the finite-size effect of lattice geometry and co-ordination number on the scaling properties of the path of the random walker.  
\end{abstract}
%---------------------------------------------------------------------------------------------------------------------------------------------------

\maketitle


%------------------------------------------------------------------------------------------------------------

%=============================================
\section{Introduction}{\label{sec:I}}
%============================================= 

The random movement of particles can nicely be analyzed by random walks where the walker is analogous to be a point-like particle moving randomly through space \cite{xia2019random,pearson1905problem,kac1947random,rubinstein2016simulation}. The space can be constructed out of  lattices and the walker can randomly move from one lattice site to another. This allows us to  obtain the  distributions of the walker’s position in space. For the classical random walk, the probability of finding the walker is higher towards the walker’s initial position. This causes the  probability distribution after the $n$-th step of a classical random walk to acquire a  binomial distribution. 
This is significantly different from that of the quantum random walk where the quantum particle tends to spread out and probability of findings the particle is always more away from the particle’s initial position \cite{kempe2003quantum,ambainis2003quantum,zhou2021review}.  This property is examined by investigating the root mean square displacement (RMSD) which essentially captures the variance of the distribution.  
Therefore, the variance of the distribution associated with classical and quantum random walks are different. The former one shows linear-$n$ behavior and the resulting particle distance goes as $n^{1/2}$ for classical random walk. For the latter case, the quantum particle travels a distance proportional to $n$ leading to the fact that quantum particle spread out quadratically faster as compared to their classical counterpart of random walk. There has been an extensive number of studies of classical random walk on two-dimensional (2D) square lattice \cite{caser1996topology, hughes1995random} while random walk on other 2D lattices of finite size are not explored much. We would like to study the effect of various lattice geometry on the classical random walk considering finite size of the underlying lattice.

There exist a plethora of studies on fractal properties of various irregular shapes \cite{peitgen1992irregular,xu1993fractals}. Coastlines of various countries and other natural boundaries are very good examples of fractals. The irregularities and variations in coastlines have been a subject of study that relied on concept beyond Euclidean geometry.  Any smooth curves can be accurately measured by the sum of the lengths of the  line segments when the lengths of the individual segments approach to zero. Interestingly, the coastlines do not follow a smooth curve and the above regular  measurement does not lead to an accurate result as the limit may not exist. Therefore, the roughness and complexity of the coastal lines can be captured by their fractal properties \cite{mandelbrot1967long,goodchild1980fractals}. The fractal dimensions for statistically self-similar phenomena have  applications in various fields such as,  astronomy \cite{caicedo2015fractal}, acoustics \cite{maragos1999fractal}, image analysis \cite{soille1996validity} apart from physics \cite{dubuc1989evaluating, roberts1996unbiased,chatterjee1992chaos}.
In the context of classical random walk, the path traced by the walker is irregular and thus finding the fractal dimension of this path can be an interesting exercise. There exist multiple prescription to compute  fractal dimension such as box dimension, characterizing the bulk volume (which is area in a 2D manifold) of the fractal, and hull dimension, estimating the circumference that encloses the volume\cite{falconer2003fractal,strogatz2018nonlinear,Mandelbrot1982,Hausdorff1919}. These  information quantify how densely the fractal occupies the space in which it lies.  Considering 2D lattices, we will explore the above quantities to examine the effect of the lattice structures on the emerging fractal dimensions.

It is noteworthy that entropy and statistical complexity measures have been employed to characterize how underlying spatial organization and interaction rules influence emergent macroscopic behavior in lattice-based dynamical systems
\cite{GaudianoRevelli2021EPJB,GaudianoRevelli2022PhysicaA,AmadoRevelliLamberti2025PRE}.
Given the above background and analytical studies on an infinite system \cite{PhysRevLett.58.2325,falconer2003fractal,alexander1982density,RammalToulouse1983,grassberger1983measuring}, we would like to emphasize the effect of  different finite size lattices, having distinct sub-lattice structures and the bond connectivity, on the classical random walk \cite{baxter2016exactly}. 
For example, honeycomb lattice has two sub-lattices and their associated bond connectivities are also different. One can find the same for Lieb lattices where not only bond connectivity pattern changes with sub-lattices but also number of connections are changed. Therefore, there exist distinct  lattice profile in terms of number of nearest neighbour and bond connectivity pattern with respect to the square lattice. Hence the  questions we address here are the following: Does RMSD or variance of the distribution depend on the lattice geometry? 
How do different lattice structure of finite sizes result in distinct fractal properties as compared to their counterparts in square lattice? Our study shows that RMSD is independent of the lattice structure rather it depends on the nature of the random walk.

This observation is surprisingly modified for mass fractal dimension  and
hull fractal dimension where both of them depend on the lattice
geometry in the regime of finite size. The average number of bonds and details of the lattice profile determine
the above dimensions. While the mean coordination number controls the overall finite-size behavior of mass and hull dimensions, additional geometric features like the lattice heterogeneity can 
introduce intriguing 
scaling of the mass and hull dimesions with the number of steps. 
Importantly, we show as the number of steps increases keeping the lattice size unaltered,
the box and hull dimensions approach their thermodynamic values towards the Brownian motion overcoming the finite size effect.

\color{black}

%=============================================
\section{Classical random walk on 2D lattice}{\label{sec:II}}
%============================================= 

 We consider the classical random walk on various 2D lattices to investigate the effect of  lattice profile \color{black}during the time evolution. We consider five different lattices, namely Square, Lieb, Honeycomb, Dice, and Kagome. These lattices have major differences in their structure which should lead to statistical non-uniformity during time evolution of the walker. In order to understand the statistical phenomena, we perform random walks on these lattices for a large number of walkers. 
Our aim is to investigate the standard deviation $r_n$ associated with the distribution of random walk after $n$-th step when walkers start from the origin at $n = 0$ for different lattices. It would be an assembly of walks at each instant on the 2D lattice. The distance $r_i=(x_i,y_i)$ of the $i$-th walker is measured with respect to the origin $r_i=\sqrt{x_i^2+y_i^2}$. The standard deviation of the walk out of $N$ number of walkers is given by
\begin{equation}
 r_{n}= \sqrt{\frac{\sum_{i=1}^N (r_i(n)-r_{\rm av}(n))^2}{N}}
 \label{eq:sd_crw}
\end{equation}
where $r_i(n)$ represents the position of the $i$-th walker after $n$-th step,  $r_{\rm av}(n)$ denotes the average position, obtained from the assembly of walkers, after $n$-th step $r_{\rm av}(n)= \sum_{i=1}^N r_i(n)/N$. We refer to Eq. (\ref{eq:sd_crw}) as RMSD of the walker without loss of generality.  
The effect of distinct lattice profile can lead to interesting observation in terms of the standard deviation of the above distribution when this exercise is performed over a large number of walker. We consider $500$ walkers for our calculation.

\color{black}

Going beyond the standard deviation, we study another  mathematical concept namely,  fractal dimension $d_f$ which is 
used to quantify the complexity or roughness of an irregular object.   Unlike traditional dimensions (1D, 2D, 3D) that describe simple geometric shapes, the fractal dimension can take non-integer values, reflecting how an object’s detail or pattern changes with scale. Fractal dimension characterizes the self-similarity and self-affinity of objects.
Fractal dimensions are often fractional, such as $1.5$, indicating that the object is more complex than a line (1D) but less complex than a plane (2D). Interestingly,  $1<d_f<2$ can represent an 
an object having infinite length but is contained within a finite area.  For example, the length of coastal lines depends on the nature of the measurement scale. If the smallest division of the scale is large compared to the intricate bends of the coastal line, the total length is found to be small. This length increases in a geometric progression as soon as it is measured with a fine scale whose smallest division is able to capture the roughness of the coastal lines.

\color{black}  In the present case,  the closed curve associated with the random walk is irregular in nature.
Therefore, the concept of fractal dimension is applicable here.

%=============================================
\section{Methodology for Calculation of fractal dimension}{\label{sec:II}}
%============================================

\subsection{Mass fractal dimension}

Usually, the volume $V$ of an object is measured by covering it with $N$ number of $d$ dimensional spheres of radius $\epsilon$, $V=N \epsilon^d$. The number of sphere $N$ is a function of the radius $r$ and this number changes as $\epsilon$ changes $N(\epsilon)\sim \epsilon^{-d}$ keeping the volume fixed. For the fractal one can obtain a relation $N(\epsilon) \sim \epsilon^{-d_f}$ with $d_f<d$. In other words,  we need $N(\epsilon)$ units of size $\epsilon$ to cover the fractal such that 
\begin{equation}
 d_f= {\rm lim}_{\epsilon \to 0} \frac{\log N(\epsilon)}{\log (1/\epsilon)}
 \label{eq:fd}
\end{equation}
Fractal dimension $d_f$ determines how $N(\epsilon)$ scales with the size $\epsilon$. To be precise, it determines the  capacity  to fill the space in terms of small cubes. This is also known as mass fractal   dimension or box dimension, in short, as it captures the packing complexity of the volume.
There exist various methods for determining the fractal dimension considering different implementation of the above relation. We below provide the box-counting method that we follow to obtain $d_f$. In addition to the fractal dimension, we also study the scaling property of the  perimeter of the fractal using the box-counting method.

In order to compute mass fractal dimension also known as box dimension  $d_f$, we need to use the bulk 
as well as  the boundary profiles of the closed curve. This allows us to investigate the effect of the  lattice profile on the 2D area of the closed curve for the random walk in various 2D lattices of finite sizes.   
We first construct the largest square whose arm is given by  the maximum span of closed curve. The closed curve of the random walker is spanned between $x_{\rm min}$ and $x_{\rm max}$ ($y_{\rm min}$ and $y_{\rm max}$) along $x$ ($y$)- directions. The arm of the square $L$ is given by ${\rm max}(\Delta x, \Delta y)$ where $\Delta x= |x_{\rm max}- x_{\rm min}|$ and $\Delta y= |y_{\rm max}- y_{\rm min}|$.    This makes sure that the entire closed curve is completely embedded inside this square. The first generation $g=1$ construction is associated with scale $s=0$. Let us say the length of the arm is $L=a$ and area of the square is $A=a^2$. The arms of this square is divided into two equal parts $a\to a/2=L$ leading to the formation of $4$ small square 
for scale $s=1$ which are associated with the second generation $g=2$. In the second generation the area of each square is $A=a^2/4$. 
In this manner we go to third generation $g=3$ when the arms of the earlier generation's squares are divided into two equal parts and eventually leading to $16$ square with the fact that  $a/2 \to a/4=L$ and $A=a^2/16$. One can find the following recursion relation  for the area  $A = a^2/2^{g-1}=a^2/2^s$ of the constituent squares and total
number of squares is $N_s=4^{g-1}=4^s$.  Interestingly,   all the higher generation squares may not enclose the closed curve always inside them. This enclosure can be complete as well as partial in nature.  The probability of capturing the closed curve inside the higher generation square decreases as the area $A$ of $n$-th generation square is proportional to $2^{n-1}$. Therefore, the number of squares occupying the closed curve $N_c$ is becoming increasingly less compared to the total number of squares $N_s$  as $s$ increases. For sake of simplicity, we refer $N_c$ as the number of occupied square and $N_e=N_s-N_c$  as the number of empty squares. We note that $N_e=0$ and $N_c=N_s$ for $(s,g)=(0,1)$. For our analysis of fractal dimension, we investigate the evolution of $N_c$ with $2^s$ as scale $s$ progresses. The mass fractal dimension $d_f$ is obtained from the slope of the straight-line fit from $\log N_c $ vs $\log (2^s)$ plot.

\subsection{Hull fractal dimension}

Having demonstrated the mass fractal dimension, one can also study the hull fractal dimension $d_h$ from the periphery/circumference i.e., boundary profile of the closed curve using the box counting method. Unlike  the above analysis, here we only focus on the number of squares $\tilde{N}_c$ capturing the boundary of the closed curve and discard the squares occupying the bulk of the closed curve. Therefore, $\tilde{N}_c < N_c$ as we do not take into consideration those bulk squares which are not exposed to the boundary of the closed curve i.e., boundary squares are only included. This allows us to investigate the effect of the  lattice profile on the 1D line profile of the closed curve for the random walk in various 2D lattices of finite size. As expected, number of empty square $\tilde{N}_e= N_s-\tilde{N}_c$ is more for the analysis of hull dimension compared to $N_e$ for the fractal dimension  The number of boundary squares is proportional to the hull dimension. As the mass fractal dimension signifies the complexity of the whole structure, hull dimension signifies the complexity of its boundary.  
We note that $\tilde{N}_e=0$ and $\tilde{N}_c=N_s$ for $(s,g)=(0,1)$. For our analysis of hull dimension, we investigate the evolution of $\tilde{N}_c$ with $2^s$ as scale $s$ progresses. The construction of squares of higher generation is already discussed previously for mass fractal dimension. We repeat the same procedure but now count the number of such boundary square.  The hull fractal dimension $d_h$ is obtained from the slope of the straight-line fit from $\log \tilde{N}_c $ vs $\log (2^s)$ plot.

%=============================================
\section{Lattice structures}{\label{sec:III}}
%============================================= 

%~~~~~~~~~~~~~~~~~~~~~~~~~~~~~~~~~~~~~~~~~~~~~~~~~~~~~~~~~~~~~~~~
%~~~~~~~~~~~~~~~~~~~~~~~~~~~~~~~~~~~~~~~~~~~~~~~~~~~~~~~~~~~~~~~~
\begin{figure}
	\subfigure{\includegraphics[width=0.48\textwidth]{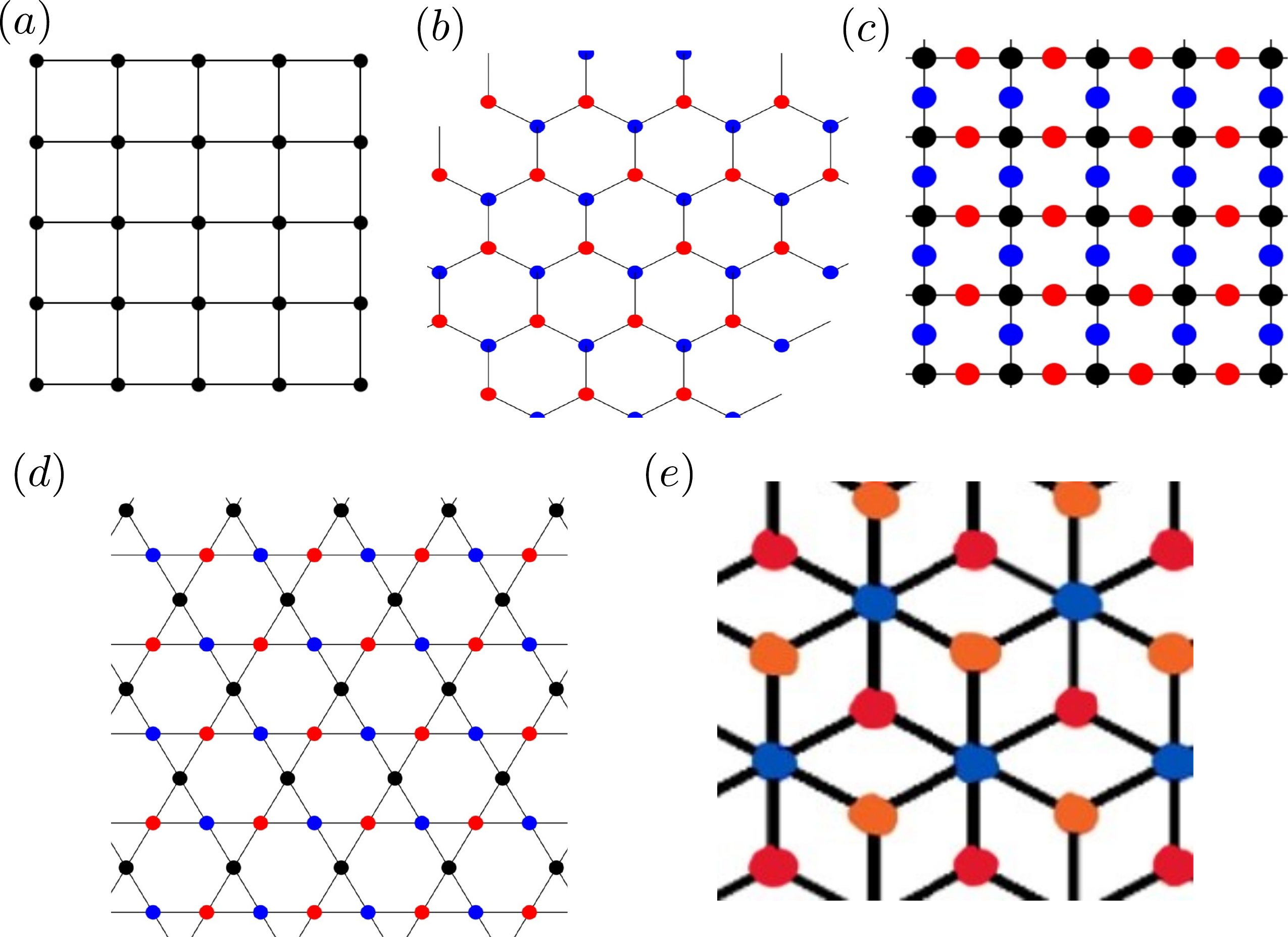}}
	\caption{(Color online) (a) Square lattice without any sub-lattice structure, (b) honeycomb lattice with two sub-lattices A (red, change it from present hollow to red) and B (blue), (c) Lieb lattice with three sub-lattices A (black), B (blue), and C (red), (d) Kagome lattice with three  sub-lattices A (black, change it from present red to black), B (blue), and C (red, change it from present  white to red), (e) dice lattice with three sub-lattices A (black, change it from present red to black), B (blue, change it from present white to blue), and C (red, change it from present orange to red). 
}
	\label{fig:lattice}
\end{figure}
%~~~~~~~~~~~~~~~~~~~~~~~~~~~~~~~~~~~~~~~~~~~~~~~~~~~~~~~~~~~~~~~~
%~~~~~~~~~~~~~~~~~~~~~~~~~~~~~~~~~~~~~~~~~~~~~~~~~~~~~~~~~~~~~~~~

 Our choice of lattice structures ensures that the walkers at every time step do not see the same lattice environment except in the case of square lattice where the walker always finds identical environment as the nearest neighbours are connected uniformly.  In the case of the different types of environment for the sites, the connectivity depends   
on the nature of the sub-lattice that the walker is instantaneously present at. In other words,  for every instance of the walker, it sees an environment that is relied on the present location of the walker. This introduces a temporal non-uniformity in the movement. To be precise, this is dimerized or trimerized ensuring a  deviation from the independent and identically distributed nature of the walk. The non-uniformity 
during the  evolution of the walker on the lattice can be introduced by various means such as having non-uniform probability distribution and non-identical connectivities over the sites. For our purposes, we have all the above ingredients. However, the connectivity is maintained by the particular choice of the lattice structure leading to a  rich profile in the time evolution of the walker.

\textit{Square lattice:} The square lattice is the simplest 2D structure where each point has four nearest neighbour with one atom per unit cell \cite{baxter2016exactly,brush1967history} having independent and identically distributed environment. The lattice has four-fold $C_4$ rotational symmetry about the perpendicular $z$ axis. All the sites are identical and shown by the hollow circle in Fig. \ref{fig:lattice} (a) leading to four probabilities $P_{\rm up}$, $P_{\rm down}$, $P_{\rm left}$ and $P_{\rm right}$ of going to the up, down, left and right into the bulk of the system. 
 The walker on the 2D square grid  is considered to  move up, down, left, or right with equal probability i.e., $P_{\rm up}=P_{\rm down}=P_{\rm left}= P_{\rm right}=0.25$ for
all the  directions.  The standard 2D random walk is symmetric, meaning that
the walker has equal probabilities of moving in each of the four directions which represent an unbiased random walk.

\subsection{Honeycomb lattice}

The honeycomb lattice is the next simplest lattice in 2D with two inequivalent atoms known as $A$ (red) and $B$ (blue) sub-lattices, per unit cell \cite{yang2018structure,novoselov2005two}, see Fig. \ref{fig:lattice} (b). This lattice has six-fold $C_6$ rotational  symmetry about the $z$ axis. The number of nearest neighbour atom is $3$ for the both these sub-lattices while their connectivity is maintained via $Y$ and inverted-$Y$ shape bonds for $A$ and $B$ type of sub-lattices, respectively. The lattice profile deviates from independent and identically distributed environment.
We consider unbiased situation i.e., probability of choosing any path within the $3$ paths forming the $Y$ or inverted-$Y$ links is the same $P_{\rm down}=P_{\rm rt}=P_{\rm lt}=1/3$ and $P_{\rm up}=P_{\rm ld}=P_{\rm rd}=1/3$ ensuring $P_A=\sum_{i \in Y-{\rm links}}P_i=1$ and $P_B=\sum_{j \in {\rm inverted}-Y-{\rm links}}P_j=1$, respectively.    
Since the random walker traverses over two types of sub-lattices, there exist a dimerized nature of the walk unlike the walk on the square lattice. To be precise, the walker sees two types of connectivity whether it is located on sub-lattice $A$ or $B$. The walker 
chooses $A$ and $B$ alternatively for its evolution. This may cause interesting outcome as far as the mass and hull fractal dimensions  are concerned.

\subsection{Lieb lattice}

Further complexifying the lattice from two sub-lattices to three sub-lattices, we consider Lieb lattice that has $C_4$ symmetry \cite{mukherjee2015observation,mielke1992exact} deviating from independent and identically distributed, see Fig. \ref{fig:lattice} (c). 
The $A$ sub-lattice (black) has '$+$'-like connectivity with blue $B$ sub-lattice connected along up and down, red $C$ sub-lattice connected along left, right directions. For red $C$ (blue $B$) lying at the middle of the horizontal '$-$'-like (vertical '$|$'-like) arm, one can find left and right (up and down) connection to the black $A$ sub-lattice. We consider unbiased walk with $i$-th link carrying the probability $P_i=1/N$ where $N$ represents the total number of link associated with the sub-lattices. 
This ensures  $P_{A}=\sum_{i \in + {\rm links}}=1/4+1/4+1/4+1/4=1 $, $P_{B}=\sum_{i \in | {\rm links}}=1/2+1/2=1$ and $P_{C}=\sum_{i \in - {\rm links}}=1/2+1/2=1$. Therefore, the walker sees a trimerized lattice during the time evolution. Unlike the previous case, if walker is at the red $C$ (blue $B$) sub-lattice, it can only move towards the black $A$ sub-lattices following the horizontal (vertical) paths. The walker can not move between $B$ and $C$ sub-lattices directly rather via the
sub-lattice $A$ only.

%~~~~~~~~~~~~~~~~~~~~~~~~~~~~~~~~~~~~~~~~~~~~~~~~~~~~~~~~~~~~~~~~
%~~~~~~~~~~~~~~~~~~~~~~~~~~~~~~~~~~~~~~~~~~~~~~~~~~~~~~~~~~~~~~~~
\begin{figure}
%	\subfigure{\includegraphics[width=0.48\textwidth]{fig1-sd.pdf}}
%\subfigure{\includegraphics[width=0.48\textwidth]{sd.png}}
\subfigure{\includegraphics[width=0.48\textwidth]{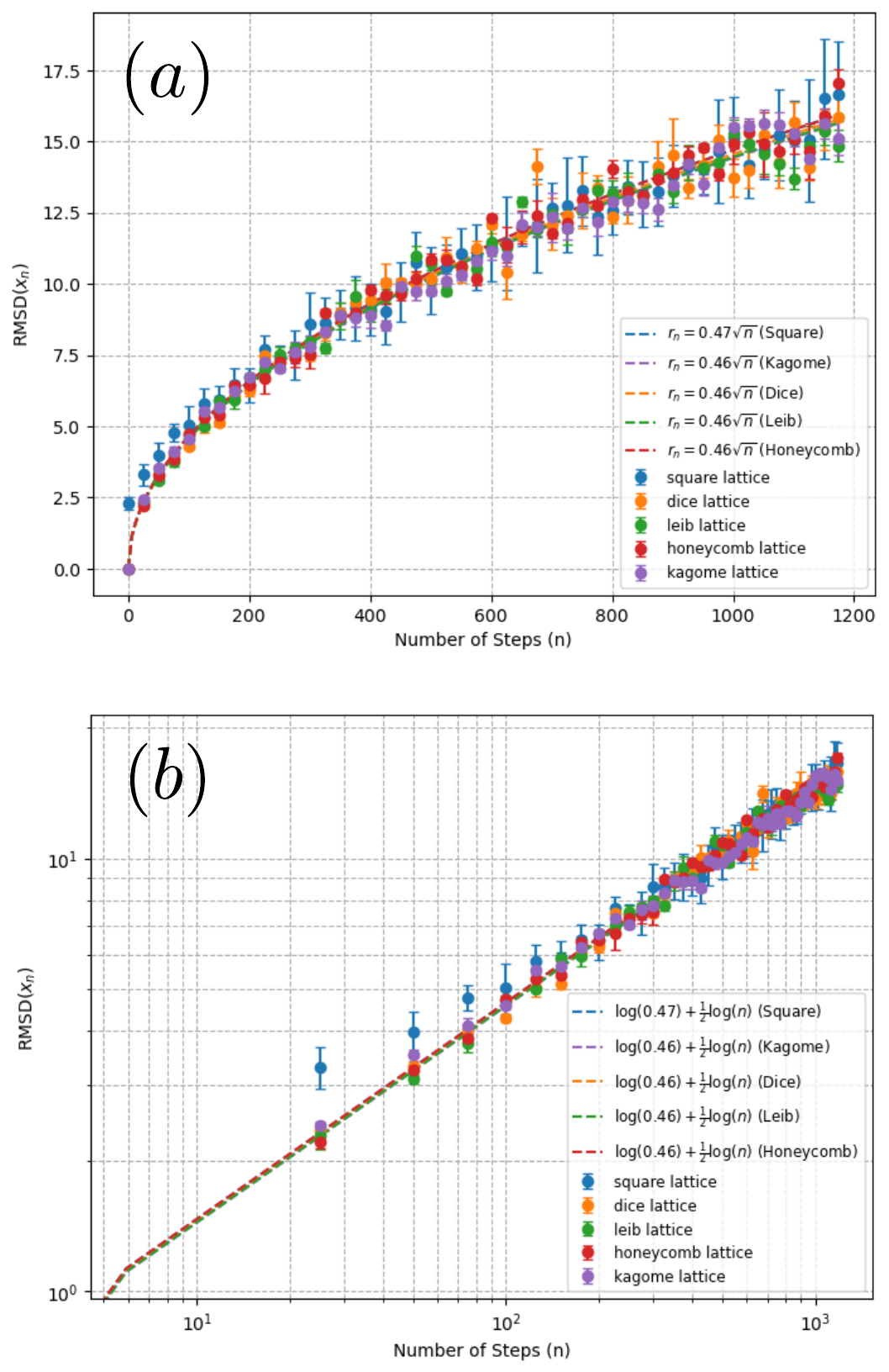}}
	\caption{(Color online) (a) We show standard deviation of displacement/ root mean square displacement (RMSD) $r_{n}$, designated by points,
	as a function of number of steps $n$ for square (violet), honeycomb (yellow), Lieb (blue), Kagome (red) and dice (green) lattice exhibiting $n^{1/2}$ fit as shown by the dashed line. (b) We show their behavior log-log scale to examine the   $n^{1/2}$ scaling more explicitly. The fitting parameters are shown for all the lattices and are in very good agreement with $n^{1/2}$ scaling.   
    We consider $500\times 500$ lattice points for all the lattices. The points represent average RMSD over  $1000$ walkers while the vertical lines associated with the points denote error bars.  
}
	\label{fig:scaling_sd}
\end{figure}
%~~~~~~~~~~~~~~~~~~~~~~~~~~~~~~~~~~~~~~~~~~~~~~~~~~~~~~~~~~~~~~~~
%~~~~~~~~~~~~~~~~~~~~~~~~~~~~~~~~~~~~~~~~~~~~~~~~~~~~~~~~~~~~~~~~

%~~~~~~~~~~~~~~~~~~~~~~~~~~~~~~~~~~~~~~~~~~~~~~~~~~~~~~~~~~~~~~~~
%~~~~~~~~~~~~~~~~~~~~~~~~~~~~~~~~~~~~~~~~~~~~~~~~~~~~~~~~~~~~~~~~
\begin{figure}[ht]
%	\subfigure{\includegraphics[width=0.48\textwidth]{fig2-fd.pdf}}
%\subfigure{\includegraphics[width=0.48\textwidth]{bulk_fixed_slope.png}}
\subfigure{\includegraphics[width=0.48\textwidth]{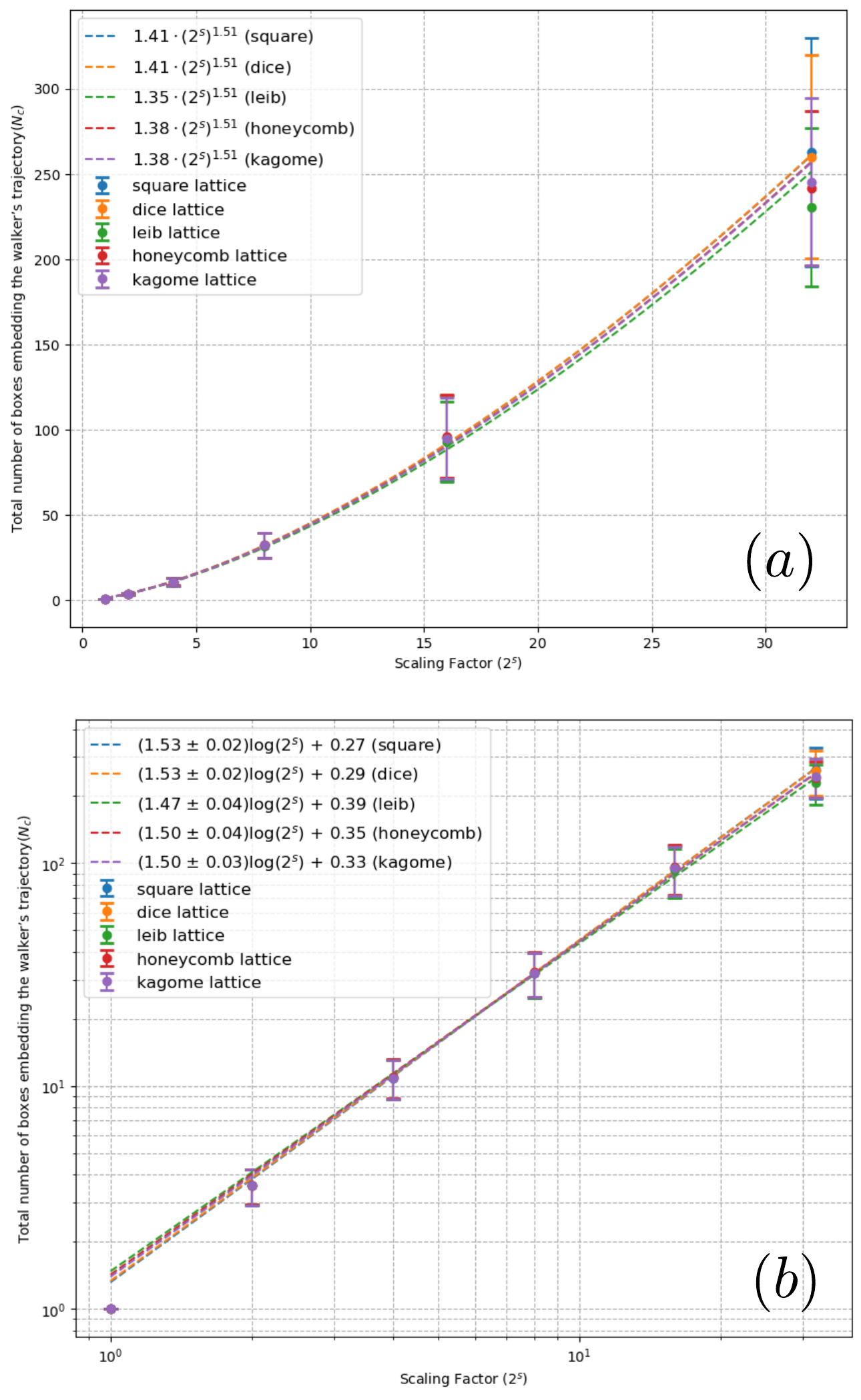}}
\caption{(Color online) (a) We show the variation of total number of squares $N_c$, designated by points, that entirely fill the area of the 2D closed curve associated with the random walk as a function of $2^s$ for square (violet), honeycomb (yellow), Lieb (blue), Kagome (red) and dice (green) where $s$ denotes the scale. We fit the data points with $\alpha(2^s)^1.51$, shown in dashed lines, by varying $\alpha$. (b) Using the straight line fit between $N_c$ and $2^s$ in log-log scale, we obtain the fractal dimension from the slopes which are different for different lattices. 
We consider $500\times 500$ lattice points for all the lattices. The walker takes $1200$ steps before we study the fractal property of the closed curve obtained from the walker.
The points represent average $N_c$ over $150$ walkers while the vertical lines associated with the points denote error bars.   
}
	\label{fig:scaling_fd}
\end{figure}
%~~~~~~~~~~~~~~~~~~~~~~~~~~~~~~~~~~~~~~~~~~~~~~~~~~~~~~~~~~~~~~~~
%~~~~~~~~~~~~~~~~~~~~~~~~~~~~~~~~~~~~~~~~~~~~~~~~~~~~~~~~~~~~~~~~

\subsection{Kagome lattice}

Continuing with the three sub-lattices 2D lattice, we consider Kagome lattice \cite{mekata2003kagome,syozi1951statistics} that deviates from independent and identically distributed, see Fig. \ref{fig:lattice} (d). The Kagome lattice has $C_6$ rotation along the $z$ axis, the inversion symmetry $I$ with respect to the center of the Kagome hexagonal center and the mirror symmetry about the $yz$ plane. 
The sub-lattice $A$ (black) is connected to sub-lattices $B$ (blue) and $C$ (red) along $\backslash$ and $/$ directions. The  blue $B$ (red $C$) sub-lattice has '$-$', and '$\backslash$'- ('$/$'-)like links connecting $C$ ($B$), and $A$ sub-lattices, respectively. Unlike the above Lieb lattice, all the sub-lattices have $4$ bonds connecting the nearest neighbours. This causes  probability for $i$-th link to be $P_i=1/4$ ensuring the unbiased nature of the  random walk. Due to the lattice geometry there exists a non-uniformity in the direction while choosing the link.

\subsection{Dice lattice}

We consider inter-penetrating honeycomb lattice where the vertex of the second hexagon (represented by red $C$ and blue $B$ sub-lattices) lies at the center of the first hexagon (represented by black $A$ and blue $B$ sub-lattices) \cite{tamang2021floquet,sutherland1986localization},  see Fig. \ref{fig:lattice} (e). This lattice also does not have independent and identically distributed.
This structure causes the vertex of the first and second hexagon $B$ to coincide with each other.    This leads to three sub-lattices $A$, $B$ and $C$ hosting inverted $Y$, star $\star$, and $Y$-like links, respectively. Therefore, $A$ and $C$ have $3$ nearest neighbour connections while $B$ has $6$. We consider unbiased random walk. Each links associated with sub-lattice $A$ and $B$ have the probabilities $P=1/3$ and links originating from $C$ have probabilities $P=1/6$. The sub-lattices $A$ and $C$ are connected via sub-lattice $B$. 
This structure of the lattice introduces a non-uniformity 
of the  lattice profile when the walker walks over the lattice.

%=============================================
\section{Results}{\label{sec:IV}}
%============================================= 

We first study the standard deviation of distribution for displacement RMSD $r_{n}$ with the number of step $n$ following Eq. (\ref{eq:sd_crw}) of the walk and examine its behavior on various lattices, see Fig. \ref{fig:scaling_sd} (a).  This analysis identifies the scaling of the  distance traveled during the walk by the walker with the number of  steps. We compute the average value of $r_n$ by the solid dots and standard deviation of the data points by the error bar. We find that $r_{n}$ for different lattices overlap with each other with the following fits: $r_n=\alpha \sqrt{n}$ with $\alpha= 0.46$ ($0.47$) for square, Kagome, and dice  (honeycomb, and Lieb) lattices. The small deviation as shown by the error bar clearly shows the validity of the $\sqrt{n}$-behavior for all the lattices signifying the universal class of diffusive transport of the random walker in 2D irrespective of the lattice environment. 
The non-uniformity of lattice profile suggests a deviation from identical and independently distributed lattice environment, however, the walk on such a lattice environment would not cause any change in the diffusive transport.  The diffusive transport does not depend on the number of the neighbour and the geometrical shape of the connecting bonds between the adjacent sites. This scaling depends only on the nature of random walk whether it is self-avoiding or not. To strengthen the above finding, we show the straight line fit  $\log r_{n}$ vs $\log n$ where the slopes are found to be $0.5$ for all lattices irrespective of their geometrical structures, see Fig.  \ref{fig:scaling_sd} (b). 
The classical random walker travels qualitatively in a similar fashion even if there are deviation from uniform lattice environment during the course of its evolution.

Having obtained universal scaling of $r_n$ for various 2D lattices of finite sizes, we further study the fractal i.e., mass and hull fractal dimensions of the walk with the aim whether the distinct  lattice profile affects the scaling behavior. We plot $ N_c $ vs $2^s$ and  $\log N_c $ vs $\log (2^s)$ in Figs. \ref{fig:scaling_fd} (a,b), respectively. We consider multiple walkers to obtain the average value, denoted by the solid dots, and standard deviation, depicted by the error bars, of $N_c$ for a given scaling factor $2^s$. We clearly find a non-linear variation of $N_c$ for all the lattices, however, the square (Lieb) lattice increases  the most (least) for larger values of the scale as obtained from their average behavior. The average values tend to overlap (deviate from) with each other for smaller (larger) values of scales. However, the error bars also increase accordingly with $s$. This makes the average values of one of the lattices to be inside the error bar associated with the other lattices. As a result, $N_c$'s are likely to scale almost identically with $2^s$ for all  the     
lattices. {We fit $N_c= \alpha (2^s)^{1.51}$ with $\alpha=1.41$, $1.38$, $1.35$, $1.38$ and $1.41$ for square, honeycomb, Lieb, Kagome and dice lattices, respectively. The numerical plots are in good agreement with the fitted parameters, except for the scale $s=5$ i.e., $2^s=32$. The interesting point to note here is that the power of $2^s$ is kept fixed for the this fit indicating to a possible  universal  scaling behavior.  } To validate this further, we examine $\log N_c $ vs $\log (2^s)$ fit with straight lines $y=mx+c$ of varying slopes $m$ and intercepts $c$, see Fig. \ref{fig:scaling_fd} (b). The solid dots almost overlap with each other, resulting in the slopes $d_f=1.53$, $1.53$, $1.50$, $1.50$, and $1.47$  for square, dice, Kagome, honeycomb, Lieb  lattices, respectively. 
The  geometrical shape of the unit cells in the lattices of finite size does not cause the mass fractal dimension $d_f$ to change substantially rather they all lie within the range of $d_f=1.50 \pm 0.03 $.

%~~~~~~~~~~~~~~~~~~~~~~~~~~~~~~~~~~~~~~~~~~~~~~~~~~~~~~~~~~~~~~~~
%~~~~~~~~~~~~~~~~~~~~~~~~~~~~~~~~~~~~~~~~~~~~~~~~~~~~~~~~~~~~~~~~
\begin{figure}[ht]
%	\subfigure{\includegraphics[width=0.48\textwidth]{fig3-hd.pdf}}
%\subfigure{\includegraphics[width=0.48\textwidth]{hull_fixed_slope.png}}
\subfigure{\includegraphics[width=0.48\textwidth]{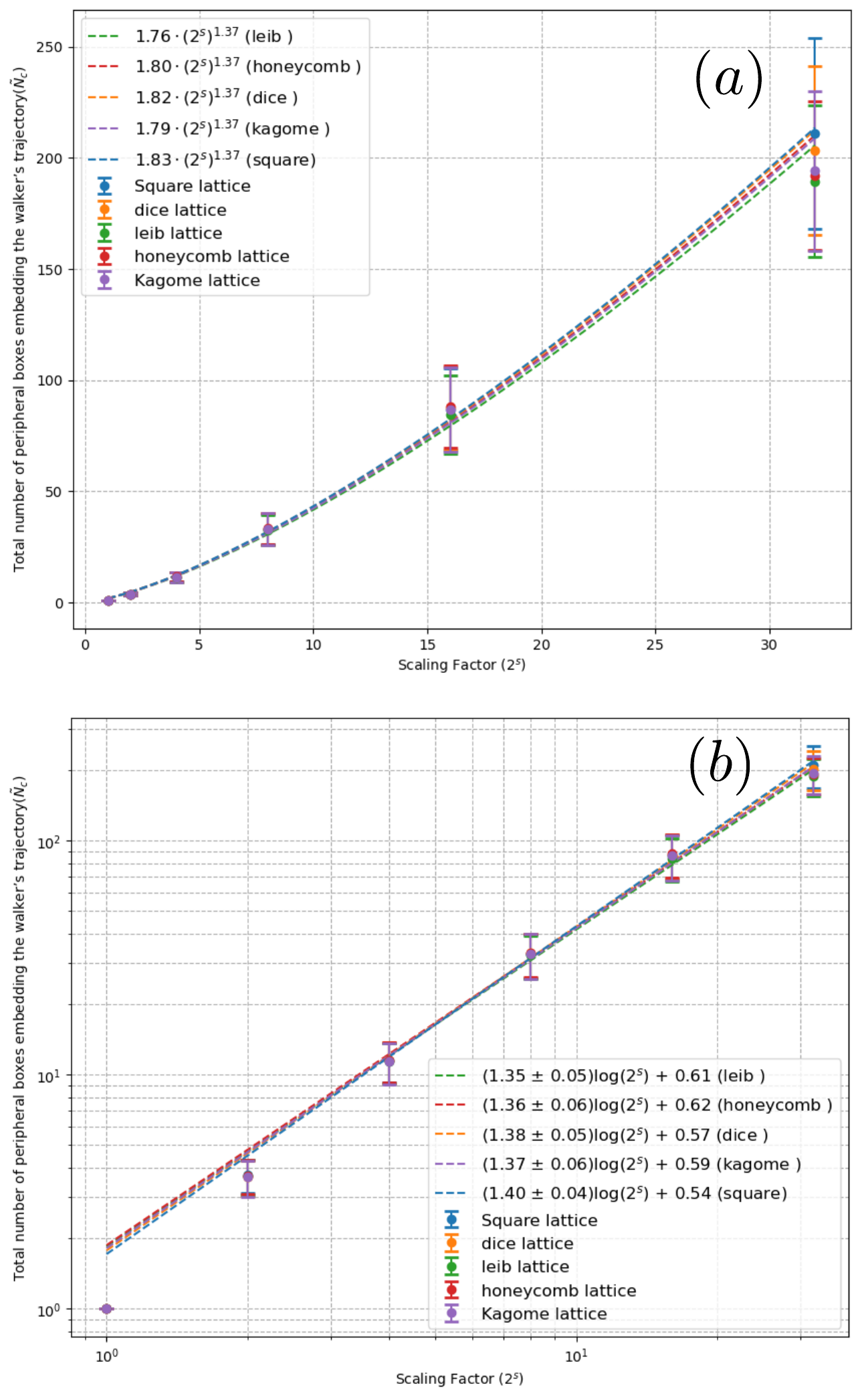}}
\caption{(Color online) 
(a) We show the variation of total number of squares $\tilde{N}_c$, designated by points, that only fill the circumference of the 2D closed curve associated with the random walk as a function of $2^s$ for square (violet), honeycomb (yellow), Lieb (blue), Kagome (red) and dice (green) where $s$ denotes the scale. We fit the data points with $\tilde{\alpha}(2^s)^1.37$, shown in dashed lines, by varying $\tilde{\alpha}$. (b) Using the straight line fit between $\tilde{N}_c$ and $2^s$ in log-log scale, we obtain the hull dimension from the slopes which are different for different lattices.  
We consider $500\times 500$ lattice points for all the lattices. The walker takes $1200$ steps before we study the fractal property of the closed curve obtained from the walker.
The points represent average $\tilde{N}_c$ over $150$ walkers while the vertical lines associated with the points denote error bars. 
}
	\label{fig:scaling_hd}
\end{figure}
%~~~~~~~~~~~~~~~~~~~~~~~~~~~~~~~~~~~~~~~~~~~~~~~~~~~~~~~~~~~~~~~~
%~~~~~~~~~~~~~~~~~~~~~~~~~~~~~~~~~~~~~~~~~~~~~~~~~~~~~~~~~~~~~~~~

The analysis on the fractal dimension further signifies that the  lattice profile does not have significant impact on the scaling of the volume of the 2D closed curve associated with an unbiased walk. The fractal dimensions for the square and dice lattice are found to be high as compared to the other lattices. This can be naively understood from the fact that dice and square lattices both have four  bonds connecting the nearest neighbours sites in average for all the steps.   The dice lattice follows $(3,3,6)$ connectivity pattern for three different sub-lattices $(A,C,B)$. Therefore, the average number of bonds over these three sub-lattices is $(3+3+6)/3=4$ which is the same as that of a square lattice without any sub-lattice structure. 
However, there is no randomness in terms of the number of connections for square lattice while for dice lattice the connectivity patterns varies on the choice of the sub-lattice.
The Lieb lattice follows $(4,2,2)$ connectivity pattern for three different sub-lattices $(A,B,C)$ leading to the lowest number of average bonds  i.e., $(4+2+2)/3=2.66$ among all the lattices. This may result in the lowest fractal dimension for the Lieb lattice.   There exist non-uniformity in terms of the number of connections in addition to the lattice profile in their connectivity patterns in all the lattices except for the square lattice. 
This induces a deviation from identically distributed lattice environment
for the walker in terms of its directionality.
%This may be the reason behind $d_f=1.605$ to be relatively small for Kagome compared to the square lattice with $d_f=1.628$ where one connectivity remains unaltered  due to the presence of a single sub-lattice. 
On the other hand, there exist three (two) and six (four) bonds connecting the nearest neighbours sites in dice (Lieb) lattice leading to a non-uniformity in terms of the number of connections in addition to the  lattice profile in their connectivity patterns. This may lead to two extreme $d_f=1.53$ ($1.47$) for dice (Lieb) lattice. For both the above lattices we have three sub-lattices similar to Kagome lattice. However, unlike the Kagome lattice the number of connection varies with sub-lattices in  dice and Lieb lattices.

Between the above two extremes of $d_f>1.5$ and $d_f<1.5$, there exist honeycomb lattice having three bonds always for two sub-lattices. In this way, there is no non-uniformity in the number of connections while the directionality of the connecting bonds changes depending on the type of sub-lattice. We obtain $d_f=1.50$ for honeycomb lattice lying between dice and Lieb lattices. We find the following relation for $d_f$ in different lattices(upto the second order): $d^{\rm S}_f=d_f^{\rm D}>d_f^{\rm HC}= d_f^{K}>d_f^{L}$ with S, K, HC, D and L stand for square, Kagome, honeycomb, dice and Lieb lattices. This order can also be naively understood from the average number of connecting bonds which is $4$ ($2.66$) for square, Kagome and dice (Lieb) lattices. In addition to the above, the variations in $d_f$ can also be caused by the  degree of the distinct nature of the  lattice  in terms of number of nearest neighbour and directionality of the connecting bonds which is the       
most for the Lieb lattice and the least for the square lattice. In short, the probability $P$ gets maximally distributed for Lieb and dice lattices while there is no distribution of $P$ in square lattice. The average number of connecting bonds is the biggest decider for the order in fractal dimension. 
Non-uniformity, quantified by the number of sublattices, causes $d_f$ to change beyond first two decimal places, see Table \ref{tab:df_ci_lattices} where we find $d^{\rm S}_f>d_f^{\rm D}> d_f^{K}>d_f^{\rm HC}>d_f^{L}$ along with their confidence intervals (CIs). \textcolor{black}{Note that the above ordering is solely based on the mean value of $d_f$ for different lattices.} This explains the equality upto the second decimal place in the $d_f$ values for honeycomb and Kagome despite Kagome having higher average number of connecting bonds. This can also be seen as the reason for the dice lattice to have slightly lower average $d_f$ than square. Average number of connecting bonds is the same for both Kagome and dice but dice has higher maximum connectivity for a given sublattice which could be the reason for $d_f^{\rm D}>d_f^{K}$.
The mass fractal dimension is able to capture the degree of the non-uniformity while the standard deviation $r_n$ is not able distinguish between different lattices of finite sizes.

Continuing further, we examine hull dimension $d_h$ where instead of counting the number of squares embedding the closed curve as a whole, we count the number of squares lying on the circumference of the closed curve. We show $ \tilde{N}_c $ vs $2^s$ and  $\log {\tilde N}_c $ vs $\log (2^s)$ in Figs. \ref{fig:scaling_hd} (a,b), respectively.
We find that $\tilde{N}_c$ increases in a non-linear fashion with $d^s$ as $s$ increases yielding $d_h>1$. We obtain that ${\tilde N}_c$ increases maximally for square lattice and minimally for the Lieb lattice, as clearly visible  for $s=5$ i.e., $2^s=32$ in Fig. \ref{fig:scaling_hd} (a). We fit ${\tilde N}_c= {\tilde \alpha} (2^s)^{1.37}$ with ${\tilde \alpha}=1.83$, $1.79$, $1.82$, $1.80$ and $1.76$ for square, Kagome, dice, honeycomb, and Lieb lattices, respectively. The numerical plots are in good agreement with the fitted parameters except for the scale $s=5$ where the curves are quite off with respect to the data points. Therefore, they tend towards the same universality class as the power of $2^s$ is kept fixed for the this fit indicating to a universality in the scaling behavior.   However, continuing with the logarithmic scale we analyze the straight line $y=mx +c$ fit for $\log {\tilde N}_c $ vs $\log (2^s)$ from which we obtain the values of slopes. These slopes correspond to the hull dimension $d_h= 1.40$, $1.37$, $1.38$, $1.36$, $1.35$ for square, Kagome, dice, honeycomb, and Lieb lattices, respectively  upto the second decimal place. Furthermore, comparing  $d_h$ and $d_f$ up to  the second decimal places, one can find that  $d_h$ sequence  matches qualitatively well with the order of  $d_f$  for different lattices except for the square (honeycomb) and dice (Kagome) as they acquire identical $d_f$. Importantly, when it comes to the comparison between $d_h$ and $d_f$ up to four decimal places, 
we find exactly the same sequence  $d^{\rm S}_h>d_h^{\rm D}>d_h^{\rm K}> d_h^{HC}>d_h^{L}$ as that of fractal dimension, see Table \ref{tab:dh_ci_lattices} along with their CIs. \textcolor{black}{Note that the above ordering is solely based on the mean value of $d_h$ for different lattices.} The exponent indicating the hull fractal dimension $d_h$ is found to be within $1.37\pm 0.03$ for all the lattices.

We qualitatively find similar trend that $d_h$ takes higher values for square and dice lattices as what is seen for $d_f$. The same reason of having four nearest neighbour irrespective of the sub-lattice can be attributed behind this observation. The Lieb  lattice exhibits the lowest $d_h$ where the average number of connecting bond is $2.66$ which is the lowest among all the lattices leading to lowest value of $d_h$ among all the lattices.  On the other hand, for dice  (Lieb) lattices, the number of nearest neighbour varies between $3$ and $6$ ($2$ and $4$) depending upon the sub-lattice type. This may result in acquiring relatively higher (lower) values of hull dimension $d_h$.   Therefore, the hull and mass fractal dimension both follow the same sequence which is mainly caused by the average number of connecting bonds and non-uniform lattice profile in terms of the number of connecting bonds and the associated directionalities. To hihglight similar qualitative effects of co-ordination number and lattice environment on the mass fractal and hull fractal dimensions, 
we show the exact mean value along with CI upto four decimal places for fractal and hull dimensions in Tables \ref{tab:df_ci_lattices} and \ref{tab:dh_ci_lattices}, respectively. The CI is determined by the standard deviation of the data points as denoted by the error bars in Figs.  \ref{fig:scaling_fd} and \ref{fig:scaling_hd}. One can find exactly the same sequence of  $d_f$ and $d_h$ across different lattices signifying the identical lattice effects in both the fractal dimensions.

%%%%%%%%%%%%%%%%%%%%%%%%%%%%%%%%%%Table

\begin{table}[h!]
\centering
\renewcommand{\arraystretch}{1.2}
\begin{tabular}{lcc}
\hline
\textbf{Lattice} & \textbf{Mean $d_f$} & \textbf{95\% CI} \\
\hline
Lieb      & $1.4722$ & $[1.4618,\; 1.4825]$ \\
Honeycomb & $1.4960$ & $[1.4926,\; 1.4995]$ \\
Dice      & $1.5290$ & $[1.4846,\; 1.5734]$ \\
Kagome    & $1.5041$ & $[1.4941,\; 1.5140]$ \\
Square    & $1.5342$ & $[1.4646,\; 1.6038]$ \\
\hline
\end{tabular}
\caption{Fractal dimension $d_f$ with lattice-specific 95\% confidence intervals.}
\label{tab:df_ci_lattices}
\end{table}

%%%%%%%%%%%%%%%%%%%%%%%%%%%%%%%%%%Table

\begin{table}[ht]
\centering
\renewcommand{\arraystretch}{1.2}
\begin{tabular}{lcc}
\hline
\textbf{Lattice} & \textbf{Mean $d_h$} & \textbf{95\% CI} \\
\hline
Lieb      & $1.3545$ & $[1.3394,\; 1.3696]$ \\
Honeycomb & $1.3584$ & $[1.3381,\; 1.3788]$ \\
Dice      & $1.3841$ & $[1.3721,\; 1.3961]$ \\
Kagome    & $1.3680$ & $[1.3510,\; 1.3849]$ \\
Square    & $1.4012$ & $[1.3849,\; 1.4175]$ \\
\hline
\end{tabular}
\caption{Hull dimension $d_h$ with lattice-specific 95\% confidence intervals.}
\label{tab:dh_ci_lattices}
\end{table}

%%%%%%%%%%%%%%%%%%%%%%%%%%%%%%%%%%Table

\textcolor{black}{Having discussed the ordering of $d_f$ and $d_h$, based on their mean values, we now revisit the order from the point of view of CIs associated with them, see Tables \ref{tab:df_ci_lattices} and \ref{tab:dh_ci_lattices}. The profile of the CIs for different lattices enables us to comment on the quantitative ordering rather than the qualitative one. A careful analysis with CIs suggests that mass fractal dimension $d_f$, obatined for square and dice, are  statistically indistinguishable. The same is true for Kagome and honeycomb lattices. On the other hand, Lieb is statistically distinct from
Kagome and honeycomb. The qualitative ordering $d^{\rm L}_f <d^{\rm HC}_f < d^{\rm K}_f<d^{\rm D}_f < d^{\rm S}_f$, based on the mean value of $d_f$, is quantitatively found to be   $d^{\rm L}_f <d^{\rm HC}_f \simeq d^{\rm K}_f<d^{\rm D}_f \simeq d^{\rm S}_f$ which is supported by the CIs. Continuing the same analysis with the CIs of hull fractal dimension $d_h$, we find that the dice and Kagome are not statistically distinguishable. The same applies to  Kagome, Honeycomb, and Lieb lattices. Therefore, qualitative ordering  $d^{\rm L}_h <d^{\rm HC}_h < d^{\rm K}_h<d^{\rm D}_h < d^{\rm S}_h$, based on the mean value of $d_h$, is quantitatively given by   $d^L_h \simeq d^{\rm HC}_h \leq d^{\rm K}_h \leq d^{\rm D}_h < d^{\rm S}_h$ which is supported by the CIs. On the other hand,
the lattices with higher average coordination numbers
tend to have slightly higher fractal dimensions. These extremes are qualitatively and quantitatively consistent as far as the mean value and  CIs are concerned. For example, Lieb lattices having  lowest coordination number, tends toward lower
values, while square/dice, having higher coordination number,  toward higher values for both the mass  and hull fractal dimensions. The above observation is based on the finite number of steps. 
}

In all of the above analysis, we exploit the finite size of the lattice by considering relatively less number of steps. Therefore, our results are
focussed on the effects of finite size of the lattices on the fractal dimensions and shed light on the  properties of the above quantities significantly away from their thermodynamic values. Our   
The fractal dimension obtained from our finite size analysis is around $1.51$ for all the lattices which is greater than Koch curve but less than Sierpinski triangle \cite{barcellos1984fractal,bannon1991fractals,darst2009curious, cannon1984fractal}. The closed curve obtained from the random walk is very similar to the quadratic von Koch curve (type 2) known as Minkowski curve for which the fractal dimension is $1.50$. This falls within the range of our observed results. Therefore, the random walker's path is comparable to the Minkowski curve.  On the other hand, the hull dimension noticed from our analysis is close to $1.37$ for different lattices.  Interestingly, the sequence of the hull dimension for different lattices follows the same sequence of fractal dimensions. This indicate the packing complexity of the bulk and envelope is almost identical as far as the walker's path is concerned. 
Note that the hull and mass fractal dimensions of 2D percolation cluster in a square lattice is $1.75$ and $1.89$ in the thermodynamic limit. The hull and mass fractal dimension obtained from random walk is significantly different from the above 2D percolation cluster. Therefore, the fractal properties associated with the random walk is different from that of the percolation cluster.

%~~~~~~~~~~~~~~~~~~~~~~~~~~~~~~~~~~~~~~~~~~~~~~~~~~~~~~~~~~~~~~~~
%~~~~~~~~~~~~~~~~~~~~~~~~~~~~~~~~~~~~~~~~~~~~~~~~~~~~~~~~~~~~~~~~

\begin{figure}[htbp]
\centering
\includegraphics[width=0.48\textwidth]{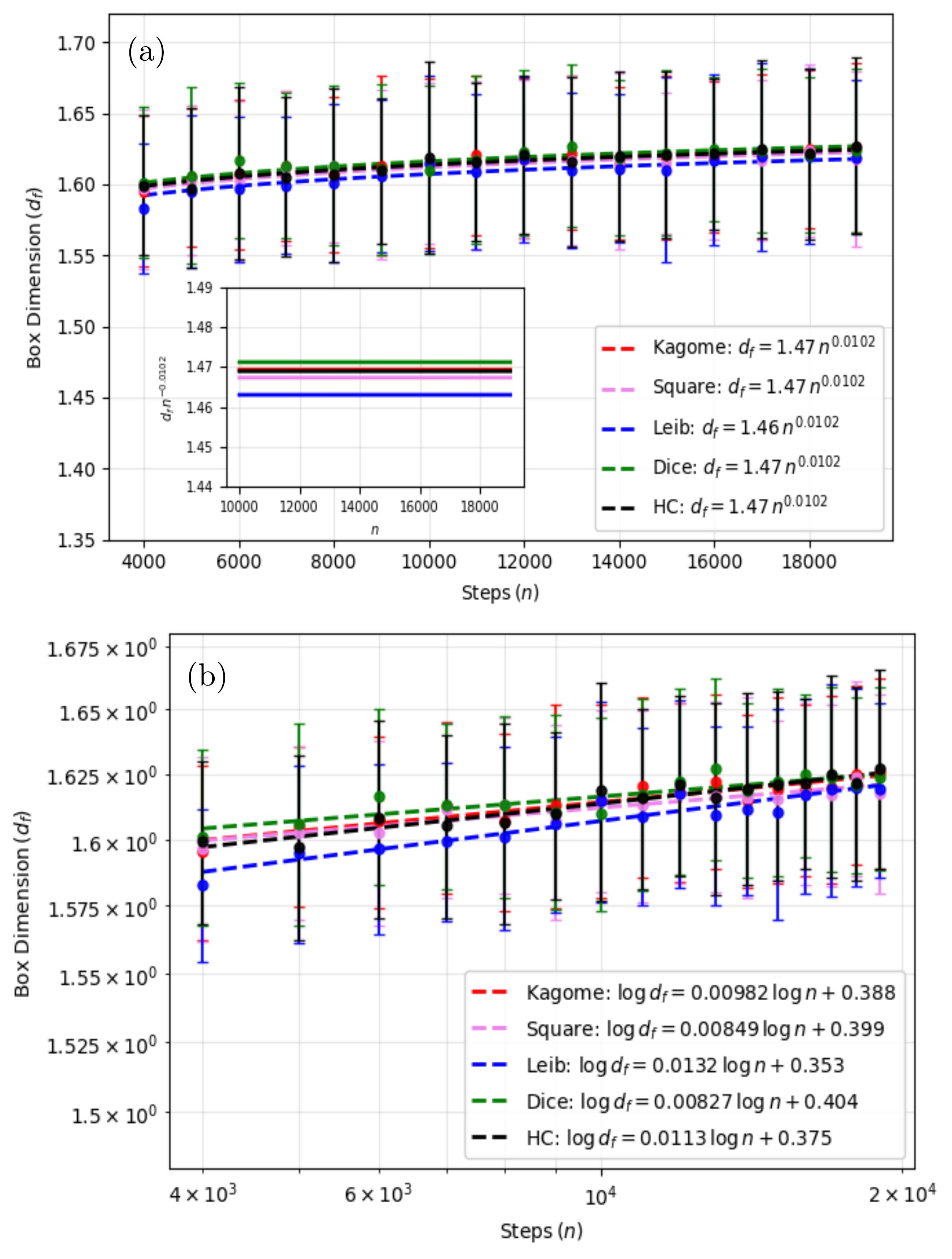}
\caption{(Color online) (a) We show the variation of mass fractal dimension $d_f$, designated by points, as a function of number of steps $n$ for square (violet), honeycomb (yellow), Lieb (blue), Kagome (red) and dice (green) lattices. We fit the data points with $\beta n^{m} $ where $m=0.0102$, shown in dashed lines, by varying $\beta$. (b) We show the variation of  $d_f$, designated by points, in log-log scale with the same colors as (a).  We fit the curves by $y=mx +c$ fit the data points with  dashed lines by varying $m$ and $c$.  We depict the data collapse $d_f n^{-0.0102}$ with steps $n$ in the insets of (a). We consider $1000\times 1000$ lattice points for all the lattices. The points represent average $d_f$ over $500$ walkers.} 
\label{fig:df_step_scaling}
\end{figure}

%~~~~~~~~~~~~~~~~~~~~~~~~~~~~~~~~~~~~~~~~~~~~~~~~~~~~~~~~~~~~~~~~
%~~~~~~~~~~~~~~~~~~~~~~~~~~~~~~~~~~~~~~~~~~~~~~~~~~~~~~~~~~~~~~~~

\begin{figure}[htbp]
  \centering
\includegraphics[width=0.48\textwidth]{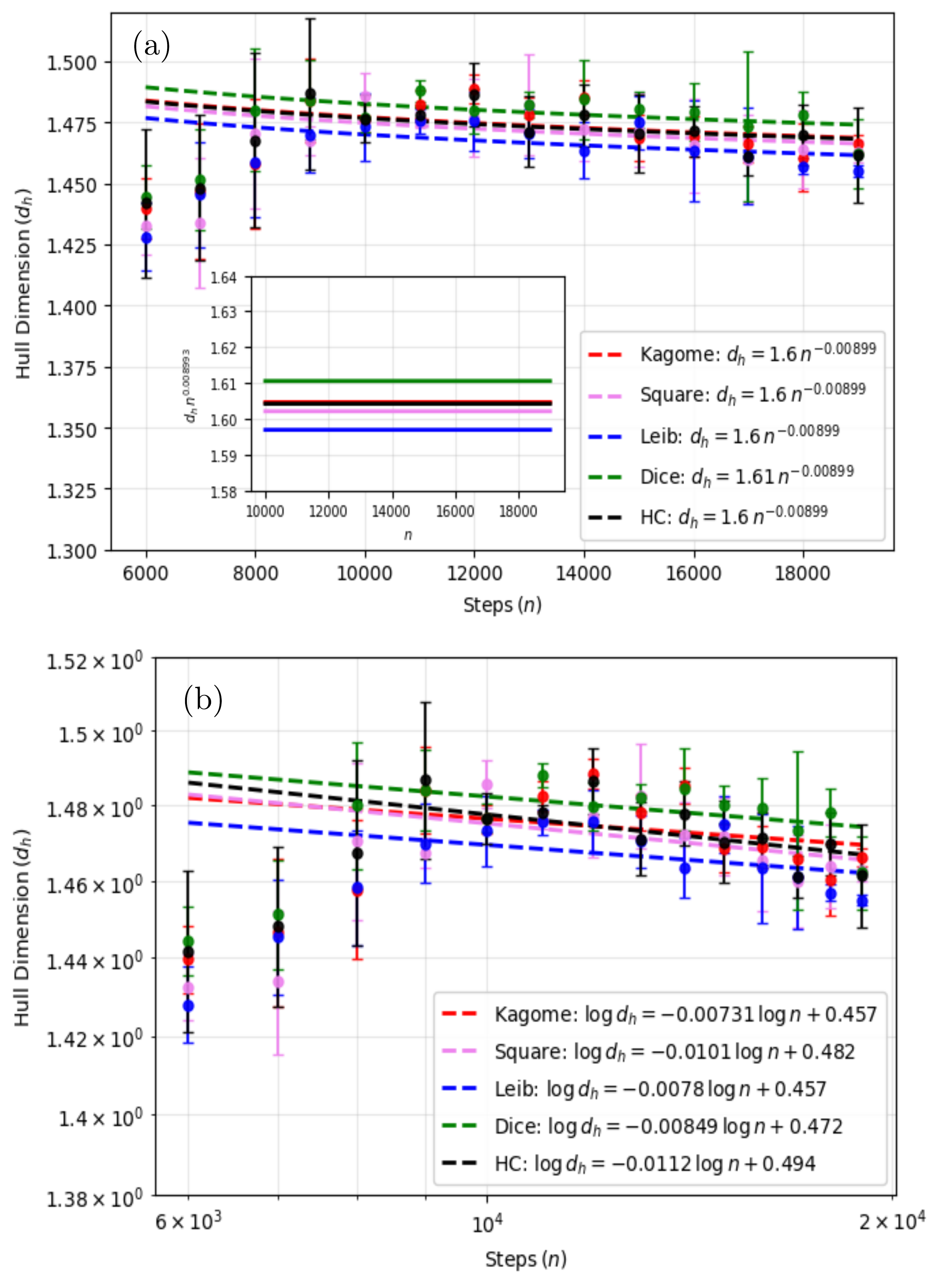}
\caption{(Color online) (a) We show the variation of hull fractal dimension $d_h$, designated by points, as a function of number of steps $n$ for square (violet), honeycomb (yellow), Lieb (blue), Kagome (red) and dice (green) lattices. We fit the data points with $\tilde{\beta} n^{m} $ where $m=0.0089$ is fixed for all lattices, shown in dashed lines, by varying $\tilde{\beta}$. (b) We show the variation of $d_h$, designated by points, in log-log scale with the same colors as (a). We fit the curves by We fit the data points with $y=\tilde{m} x+ \tilde{c} $, shown in dashed lines, by varying $m$ and $c$.  We depict the data collapse $d_h n^{-0.0089}$ with steps $n$ in the insets of (a).
We consider $1000\times 1000$ lattice points for all the lattices. The points represent average $d_h$ over $500$ walkers.}
\label{fig:dh_step_scaling}
\end{figure}

Having examined the finite size behavior of mass and hull fraction dimension, we now study the tendency to approach the thermodynamic limit upon increasing the number of steps on a relatively large lattice $1000\times 1000$. This system size allows us to increase the number of steps by one order of magnitude such that the thermodynamic behavior can be analyzed more extensively. We demonstrate the variation of  $d_f$ and $d_h$ with the number of steps $n$ from $4\times 10^3$ to $18\times 10^3$ and $6\times 10^3$ to $18\times 10^3$ in Figs. \ref{fig:df_step_scaling} and \ref{fig:dh_step_scaling}. We observe that the values of $d_f$ keep increasing very slowly with the number of steps after the initial speedy climb.
We find $d_f$ increases with $n$ while their relative order follow the almost the same trend with $n$ as the non-linear fit of $d_h$ for dice (Lieb) stays at the top (bottom). Importantly, these non-linear fit show parallel nature for different lattices i.e., $d_f=\beta n^{0.0102}$ signifying the fact that the tendency to reach the thermodynamic limit is independent of the lattice profile, see Fig. \ref{fig:df_step_scaling}(a). 
In order to emphasize the scaling with number of steps more clearly, we fit using straight line $y=mx +c$ in the log-log scale where the slopes $m$ appear to be $.0108\pm0.0025$ for all the lattices, see    Fig. \ref{fig:df_step_scaling}(b).   In Figs. \ref{fig:df_step_scaling}(a,b), we show the mean and  standard deviation of the data with filled circle and the vertical bar, respectively. Importantly, the $d_f$ profiles of the all the lattices lie inside the combined error bar window indicating towards their universal thermodynamic behavior. We show the data collapse $d_f n^{-0.0102}$ with steps $n$ in the insets of  Fig. \ref{fig:df_step_scaling}(a) where we find $n$-independent behavior. This clearly suggests the $n$-scaling of the mass fractal dimension.

Coming to the variation of $d_h$ with $n$,
we find speedy climb  and a gradual increase to about $8000$ steps but gradually decrease afterwards to a saturation value for all the lattice types. Therefore, 
we observe an exactly opposite trend for hull fractal dimension $d_h$ that decreases with $n$ when $n>10^4$  for all lattices, see Fig. \ref{fig:dh_step_scaling} (a). Importantly, for smaller $n<10^4$, the finite size effect is severe as $d_h$ increases with $n$. Importantly, $d_h$ of all lattices vary parallelly with the non-linear fit $d_h =\tilde{\beta} n^{-0.0089}$ for $n$ above $10^4$.  To examine this further in Fig. \ref{fig:dh_step_scaling} (b), we perform straight line fit $y=\tilde{m}x +\tilde{c}$ over log-log scale with $\tilde{m}$ lying between $-(0.009\pm 0.002)$ for all the lattices.   In Figs. \ref{fig:dh_step_scaling}(a,b), we show the mean and  standard deviation of the data with filled circle and the vertical bar, respectively. Similar to the $d_f$ profiles in Fig. \ref{fig:df_step_scaling}(a), we find that the $d_h$ profiles of the all the lattices lie inside the combined error bar window indicating towards their universal thermodynamic behavior. We show the data collapse $d_h n^{-0.0089}$ with steps $n$ in the insets of  Fig. \ref{fig:dh_step_scaling}(a). The  $n$-independent profile of the data collapse  clearly suggests qualitatively identical scaling of hull fractal dimension to that of  mass fractal dimension in thermodynamic limit.  \textcolor{black}{This data collapse of mass and hull dimensions can be attributed to the Brownian motion in 2D in the thermodynamic limit which is independent of the lattice-specific distinctions.} Therefore, the scaling of $d_f$ and $d_h$ show qualitatively similar scaling exponent $\alpha$ with $n^{\alpha}$ and $n^{-\alpha}$ profile, respectively.   
Interestingly, the relative order of $d_h$ and $d_f$ for 
different lattices remains identical indicating to the fact that they all belong to the same universality class in the thermodynamic limit.

%~~~~~~~~~~~~~~~~~~~~~~~~~~~~~~~~~~~~~~~~~~~~~~~~~~~~~~~~~~~~~~~~
%~~~~~~~~~~~~~~~~~~~~~~~~~~~~~~~~~~~~~~~~~~~~~~~~~~~~~~~~~~~~~~~~

%~~~~~~~~~~~~~~~~~~~~~~~~~~~~~~~~~~~~~~~~~~~~~~~~~~~~~~~~~~~~~~~~
%~~~~~~~~~~~~~~~~~~~~~~~~~~~~~~~~~~~~~~~~~~~~~~~~~~~~~~~~~~~~~~~~

We now comment on the possible thermodynamic limit of our findings. The increasing nature of hull fractal dimension $d_h$ and mass fractal dimension $d_f$ with number of steps  $n$ indicates the underlying connection with the Brownian motion. It is known that the 
mass fractal dimension of the Brownian motion in 2D is found to be $2$\cite{falconer2003fractal} while the hull fractal dimension becomes $4/3$ \cite{LawlerSchrammWerner2001}. In the present case, $d_h$ and $d_f$ may asymptotically approach the Brownian limit, however, these fractal dimensions are significantly different when we are away from the thermodynamic limit. 
According to Donsker’s invariance principle\cite{donsker1951,kipnis1986,ethier1986}, the distribution of a random walk taking place on 
an independent and identically distributed converges to that of Brownian motion under appropriate scaling.  In particular, under the diffusive scaling with the space being normalized by \(\sqrt{n}\), the rescaled distribution converges to a standard Brownian motion. Interestingly, adopting the same idea in our context, one can connect the random motion in different lattices to an underlying Brownian motion. The independent and identically distributed is only noticed for square lattice while  the random walks on structured lattices 
such as the Lieb or honeycomb lattice are not identically 
distributed since the step distribution depends on the 
sub-lattice type.
Given the fact that all the lattices  follow qualitatively and quantitatively the same trend as that of the  square lattice, one can comment that in the thermodynamic limit lattice structure becomes irrelevant and the fractal dimensions converge to that of a  Brownian motion in 2D. However, our study numerically reveals the finite size behavior of fractal dimensions where analytical treatments do not work.

It is important to note that critical phenomena for non-interacting and interacting models of statistical physics can be explained by the regular random and self-avoiding random walk, respectively \cite{kipnis1986,ethier1986,nienhuis1982,LawlerSchrammWerner2004, Hattori1987,Duplantier1989,RammalToulouse1983}. It has been shown that critical clusters (Ising, percolation, self-avoiding random walk) are fractal objects, often described by a single fractal dimension or even a multifractal spectrum\cite{Kadanoff1966,Wilson1971,Mandelbrot1982,Duplantier1990,EversMirlin2008}. Exploiting the connection with the critical phenomena,  the critical exponent $\nu$ associated with correlation length is found to be $1/2$ and $3/4$ for regular and self-avoiding random walks, respectively. Therefore, the mass fractal dimension is found to be $d_f=1/\nu = 2$ for regular random walk and $4/3$ for self-avoiding random walk. Given the RMSD behavior, one can comment that our results indicate towards the Brownian motion under diffusive scaling. Our findings in the thermodynamic limit indeed refer to the critical phenomena associated with mean-field Ising universality class. 
Interestingly, the fractal dimensions deviate significantly from their thermodynamic values as the independent and identically distributed lattice environment is not perceived with a finite size of the lattice. Therefore, our work is important in regards to the evolution of  
fractal dimensions with number of steps and how the above quantities approach their thermodynamic values. The finite size fractal dimensions are useful for various numerical studies with finite systems close to criticality where analytical treatments are limited.

%=============================================
\section{Conclusions}{\label{sec:V}}
%============================================= 

We consider five different lattices namely, square, honeycomb, dice, Kagome and Lieb lattices to study the effect of connectivity profiles on the two-dimensional classical random walk without self avoiding. Interestingly, these lattice structures have different number of bonds connecting with distinct types of sub-lattices leading to the walker to experience versatile lattice profile of finite sizes while executing the random walk. We first examine the RMSD of the walker where different lattices conceive the same linear-$n$ behavior irrespective of their structure. This is caused by dimensionality of the problem and non-self-avoiding nature of the walker.  We next compute the fractal dimension of the graph, capturing the bulk area of the 2D graph,  acquired by the walker to study the finite size effect of lattice structure on the fractal properties.

\textcolor{black}{We remarkably find that the different  lattice profile of the finite size lattices result in a qualitatively distinct fractal dimension as far as their mean values are concerned while the CI can rectify the distinctiveness.} We  observe that square (Lieb) lattice exhibits highest (lowest) fractal dimension as the former (later) has no (maximum) non-uniformity in their connectivity pattern. In addition to this, the mass fractal dimension is likely to be higher (lower) when more (less) number of bonds are connected to the lattice sites in an average manner for finite size lattices. \textcolor{black}{
This allows us to  understand the qualitative sequence of the values associated with fractal dimensions for these lattices while their quantitative values as confirmed by the respective CIs indeed reflect substantial statistical overlaps. For example, Kagome and honeycomb yield statistically indistinguishable mass dimension. Similarly, square and dice  have substantial statistical overlap in mass dimension. }

This same analogy holds true for hull fractal dimension, capturing the boundary circumference of the 2D graph, where the
non-uniformity of the lattice profile and average number of connecting bonds play crucial role. We analyze the hull fractal dimension and show the sequence is qualitatively the same as that of the mass fractal dimension leading to the fact that lattice structure of finite sizes indeed affects the emerging fractal nature of the graph traced by the walker. 
\textcolor{black}{Similar to the mass dimension, the qualitative trend in the sequence  of the hull dimension is quantitatively corrected after taking into account the corresponding CIs. To be precise, only square is clearly separated from the lower group where dice, Kagome,
honeycomb, and Lieb show extensive statistical overlap. However, Kagome and  honeycomb lattices share nearby values of hull dimension while Lieb can be naively considered to display the lowest hull dimension. }

We examine the  evolution of mass and hull fractal dimensions with number of steps 
to investigate the tendency to approach the thermodynamic limit which is given by Brownian motion in $2D$ with mass fractal dimesnion equal to $2$ \cite{falconer2003fractal} and hull fractal dimension equal to $4/3$ \cite{LawlerSchrammWerner2001} in $2D$. \textcolor{black} {Our analysis on data collapse with the number of steps clearly demonstrates universal scaling  toward the thermodynamic
limit, which is a strong result independent of the lattice-specific distinctions.}

 \color{black}
%======================================================
\section*{Acknowledgments}
%======================================================
T.N. acknowledges Subhadeep Roy for useful discussions on fractal and hull dimensions. N.S. and T.N. thank the
NFSG “NFSG/HYD/2023/H0911” from BITS Pilani.

{\bf Data availability statement:} The data may be available on request to the corresponding
author.

{\bf Conflict of interest statement:} We declare that this manuscript is free from any conflict
of interest. The authors have no financial or proprietary interests in any material discussed in
this article.

{\bf Funding statement:} No funding was received particularly to support this work.

{\bf Authors’ contributions:} Tanay Nag conceived the idea, analyzed the results and wrote the manuscript. Nimish Sharma did all the numerical calculations, prepared the figures, analyzed the results, and partially wrote the manuscript.

%\bibliography{bibfile1}{}


\end{document}